# GENERALIZED GALILEI-INVARIANT CLASSICAL MECHANICS*

HARRY W. WOODCOCK
*School of Science and Health*
*Philadelphia University*
*Philadelphia, PA 19144, USA*
WoodcockH@PhilaU.edu
*and*
*Department of Physics*
*Temple University*
*Philadelphia, PA 19122, USA*
*and*
PETER HAVAS (deceased)
*Department of Physics*
*Temple University*
*Philadelphia, PA 19122, USA*

To describe the "slow" motions of n interacting mass points, we give the most general 4-d non-instantaneous, non-particle symmetric Galilei-invariant variational principle. It involves two-body invariants constructed from particle 4-positions and 4-velocities of the proper orthochronous inhomogeneous Galilei group. The resulting 4-d equations of motion and multiple-time conserved quantities involve integrals over the world lines of the other n-1 interacting particles. For a particular time-asymmetric retarded (advanced) interaction, we show the vanishing of all integrals over world-lines in the ten standard 4-d multiple-time conserved quantities, thus yielding a Newtonian-like initial value problem. This interaction gives 3-d non-instantaneous, non-particle symmetric, coupled non-linear second-order delay-differential equations of motion that involve only *algebraic* combinations of non-simultaneous particle positions, velocities, and accelerations. The ten 3-d non-instantaneous, non-particle symmetric conserved quantities involve only algebraic combinations of non-simultaneous particle positions and velocities. A two-body example with a generalized Newtonian gravity is provided. We suggest that this formalism might be useful as an alternative slow-motion mechanics for astrophysical applications.

## 1. Introduction

In order to describe the motion of n interacting mass points, most "slow-motion" attempts start from the generally covariant field equations of General Relativity Theory (GRT). One approach, the Lorentz-Droste-EIH equations,[1] leads to single-time equations of motion that follow from a single-time Lagrangian which is approximately[2] Poincaré-invariant and from which, approximately Poincaré-invariant conserved quantities can be calculated by Noether's[2,3,4] theorem.

A second approach by Havas and Goldberg[5] is a "fast-motion" (exactly) Poincaré-invariant approximation yielding multiple-time integro-differential equations of motion. Their time-symmetric equations of motion follow from a Poincaré-invariant variational principle from which a single-time approximately Poincaré-invariant Lagrangian and slow-motion equations of motion can be found by expansion[6] of the exact expressions in powers of v/c.

Yet a third approach is the single-time "post-Newtonian" formalism initiated by Nordtvedt and Will[7] for metric theories of gravity. For GRT itself, Damour, Soffel, and Xu have made a thorough analysis[8] of such equations, which have been proposed[9] for use by the International Astronomical Union.

All of these single-time descriptions have Newton's instantaneous Galilei-invariant mechanics as a limit. Of course, this is not surprising since Einstein required such a limit[10] for GRT and Newton's laws of motion were successful in describing slow motions in the solar system.

But, when instantaneous Newtonian equations of motion are applied to describe mass points separated by astrophysical distances, difficulties have arisen; e.g., the measured rotation curves of spiral galaxies conflict with Newtonian predictions leading people to postulate dark matter,[11] while the measured

*Research supported in the past by the National Science Foundation.

accelerations of galaxies have been used to infer dark energy.[12] "Difficulties" should not be surprising since, by the special theory of relativity, the time delay of an interaction between two particles is expected to be the distance between particles divided by the speed of light in empty space. Interaction time delays measured in millions of *years* are embarrassingly non-instantaneous.

As an alternative to instantaneous slow-motion descriptions, we propose to start from multiple-time, non-particle symmetric Galilei-invariant 4-d equations of motion that follow[13] from a generalized non-particle symmetric explicitly Galilei-invariant variational principle. This yields ten multiple-time conserved quantities via Noether's[14,15] theorem. The resulting Galilei-invariant equations of motion and conserved quantities are, in general, not only multiple time and non-particle symmetric but involve integrals over the entire world lines of the other interacting particles, just as in Poincaré-invariant theories. Little is known about their initial value theorem.

Here, for one particular time-asymmetric interaction covering a large class of Galilei-invariant two-body potential energies (including a generalized Newtonian gravity), the Lagrangian equations of motion and corresponding conserved quantities are shown not to involve integrals over the other n-1 particles' world lines, provided that the interaction times and points of evaluation are chosen appropriately. Then, the 3-d conserved quantities involve only *algebraic* forms of non-simultaneous particle positions and velocities, similar to Newtonian theory. Hence, their 3-d equations of motion, which involve non-simultaneous position-, velocity-, and acceleration-dependent forces, have a Newtonian-like initial value problem requiring 6n non-simultaneous initial data that must be chosen to match the arbitrary but fixed interlocking set of interaction time delays. For just one pair of particles, the Galilean time delay in this formulation is not constrained, while for n>2 particles they must be interlocking.

While such a formulation is interesting in an academic sense, the one case for which it might have a useful application is as an alternative slow motion but non-simultaneous description of astrophysically-separated, gravitationally-interacting mass points. Rather than using the customary instantaneous Newtonian description for far-separated particles, we suggest choosing the *average* interparticle separation divided by the speed of light for the *constant* value of the Galilean time delay between two interacting particles. The resulting description remains Galilei invariant but thereby takes some account of the special relativistically imposed finite speed of interactions. This would be useful only for particles far away from one another, i.e., having time delays large compared to solar system time delays. Of course, if the Galilean time delay is set equal to zero (or $c \to \infty$), these Lagrangian equations of motion and conserved quantities have instantaneous Newtonian dynamics as their limit.

However, a possible disadvantage of these equations is that certain motions in generalized gravitational cases that naively one might expect to follow, do not. E.g., two equal masses gravitationally attracted toward one another along one line at equal speeds at different times cannot continue at equal speeds. Furthermore, two equal masses heading oppositely along *different* parallel lines at equal speeds but situated on a line perpendicular to both velocities do not go in a circle. General solutions to these coupled, non-linear, multiple-time, second-order, delay-differential equations are unknown. While examples like this are counter-intuitive to those who were raised on Newtonian mechanics, so also are postulations of dark matter and dark energy. Perhaps, when numerically applied to stars at the fringe of a galaxy or to a group of galaxies, this formulation might turn out to be a better description than dark matter.

An investigation of Galilei-invariant generalizations of Lagrangian forms of classical mechanics first of all requires the derivation of the two-body invariants of the proper orthochronous inhomogeneous Galilei group. Possible generalizations of Newtonian mechanics allowing for more general forces, but maintaining the conservation laws, were considered long before the connection of these laws with invariance properties of the equations was realized. Velocity-dependent terms in the Lagrangian seem to have been considered first by Schütz;[16] derivatives of arbitrary order were considered by Königsberger[17] and many-body forces by Fichner.[18] In all of these generalizations, the forces were still assumed to depend on the particle variables taken at the *same* instant of time. Simultaneity was still maintained in a later paper[19] that took invariance under the Galilei group as its starting point for the investigation of possible forms of two-body interaction.

This restriction of simultaneity was due to the fact that the invariants of the Galilei group were then unknown. Although Felix Klein clearly realized the difficulty of the problem of determining all these invariants (in contrast to the corresponding problem for the inhomogeneous Lorentz, i.e., Poincaré group) and posed it as a challenge,[20] almost everyone seems to have taken it for granted that these invariants would have to conform to the restriction mentioned. As far as we know, only one attempt early in the 20[th] century exists to determine all the two-particle invariants of the *inhomogeneous*[21] Galilei group, a direct response to Klein's challenge by Weitzenböck.[22] Unfortunately, not only are Weitzenböck's methods quite unfamiliar to physicists, and by now outdated, but also his results are incomplete and partly incorrect.

Although the 3-d forms of the Galilei invariants were published by the second author,[13] no proof was offered. In Sec. 2, we summarize for the convenience of the reader the relevant parts of Ref. 13 (FN) and we derive the two-particle invariants using, for the first time[21] in 4-d, a non-singular two-index tensor; they are presented both in 3+1 form [FN (37)] and in explicit tensor forms not previously known. A Galilei-invariant 4-d generalization of Lagrange's mechanics [FN (82N)] (with only one time-symmetric Green's function) was included in that review, but neither examples nor multiple-time conserved quantities were given. In Sec. 3, we present the most general particle-non-symmetric Galilei-invariant Lagrange interactions and their equations of motion. The ten corresponding multiple-time conservation laws[14, 15, 23] are considered in Sec. 4. Integrals over world lines in the ten Lagrangian multiple-time conserved quantities are shown to be zero for time-asymmetric retarded (advanced) interactions, but only if the time delays and points of evaluations are chosen to be appropriately interlocking. This yields a Galilei-invariant Lagrangian mechanics that has a Newtonian-like initial value problem. In Sec. 5, we give an example of this generalized mechanics. For this time-asymmetric Lagrangian example we chose the most obvious invariant for the argument of the two-body potential energy. This case is not solved, but inferences for special cases of generalized gravity already noted above are treated. The results are discussed in Sec. 6.

**2. The Galilei Group and its Invariants**

We consider the group of linear transformations of the Cartesian space coordinates $x^1, x^2, x^3$ and the time t = $x^0$

$$x'^{\mu} = \alpha^{\mu}{}_{\rho} x^{\rho} + \xi^{\mu}. \tag{1}$$

Here and in the following, Greek indices run from 0 to 3, Roman indices from 1 to 3, and summation over repeated indices is understood. The full inhomogeneous Galilei group is defined as the group of transformations (1), restricted[20] by the conditions

$$\alpha^0{}_0 = 1, \quad \alpha^0{}_r = 0, \quad \alpha^m{}_r \alpha^n{}_r = \delta_m{}^n, \tag{2}$$

where $\delta_m{}^n$ is the Kronecker delta. These conditions are nontensorial; however, for our purposes it is more convenient to work with tensorial conditions. We briefly summarize this formalism, which is described in detail in FN, Sec. III.

The conditions (2) can be shown to be equivalent to

$$g_{\mu\nu} \alpha^{\mu}{}_{\rho} \alpha^{\nu}{}_{\sigma} = g_{\rho\sigma}, \quad h^{\mu\nu} \alpha^{\rho}{}_{\mu} \alpha^{\sigma}{}_{\nu} = h^{\rho\sigma}, \tag{3}$$

where $g_{\mu\nu}$ and $h^{\rho\sigma}$ are tensors which have numerically the same components in all coordinate systems connected by (1) with (2); these components are chosen such that

$$g_{00} = 1, \quad h^{11} = h^{22} = h^{33} = -1, \tag{4}$$

and all other components vanish. Thus, these tensors are singular and satisfy

$$g_{\mu\rho} h^{\rho\sigma} = 0. \tag{5}$$

These equations imply (2) as well as

$$D \equiv \left| \partial x'^{\rho} / \partial x^{\mu} \right| = \pm 1 \tag{6}$$

for the Jacobian of the transformation. In complete analogy to the well-known structure of the Poincaré group[24], the Galilei group consists of four parts, corresponding to the four combinations of the signs of *D* and $\alpha^0{}_0$. The part with *D* = $\alpha^0{}_0$ =1 forms a subgroup, the proper orthochronous Galilei group. The physical requirement of invariance under this subgroup is more basic than that under the full group, because it is this subgroup that corresponds to uniform relative motion of frames of reference without reflections and time reversal, i.e. which expresses the equivalence of all inertial systems. If we restrict ourselves to orthochronous Galilei transformations, the quantity

$$w_\mu = (1,0,0,0) \tag{7}$$

transforms like a covariant vector, with numerically the same components in all coordinate systems; under antichronous transformations (for which $\alpha^0{}_0 = -1$), it changes sign. We have

$$h^{\mu\rho} w_\rho = 0, \quad g_{\mu\nu} = w_\mu w_\nu. \tag{8}$$

The coordinates $x^\rho$ can be considered as those of a space with affine connection, where all affine connections $\Gamma^\rho{}_{\mu\nu}$ vanish for the class of coordinates[25] used in Eq. (1). The tensor $g_{\mu\nu}$ might be used as a metric, thus defining a four-dimensional distance by the scalar

$$\Delta s \equiv (g_{\mu\nu} \Delta x^\mu \Delta x^\nu)^{\frac{1}{2}}; \tag{9}$$

however, this metric is singular and thus the space is not Riemannian. The separation (9) actually refers to a pure time interval because of the form (4) of $g_{\mu\nu}$ and assigns a separation of zero to any two simultaneous events.

Although $\Delta s$ is not suitable for characterizing a four-dimensional separation unambiguously, nevertheless it is an invariant, indeed the most important invariant of the Galilei group, expressing the fact that time differences are the same in all inertial systems—the remnant of Newton's "absolute time"—since, from the first of Eqs. (4)

$$\Delta s = \Delta x^0 = \Delta x'^0 = \Delta t = \Delta t'. \tag{10}$$

We now consider the world line of particle i whose four coordinates we denote by $z_i^\mu(\tau_i) = [t_i(\tau_i), \vec{r}_i(\tau_i)]$, which can be expressed as functions of an arbitrary Galilei-invariant parameter. It is convenient to take as this parameter the "proper time" $\tau_i$ defined by

$$d\tau_i = (g_{\mu\nu} dz_i^\mu dz_i^\nu)^{\frac{1}{2}}. \tag{11}$$

Clearly $\tau_i$ differs from $z_i^0 = \pm t_i$ at most by an additive constant[26] $C_i$. Nevertheless, it is convenient to distinguish the two quantities notationally, as it serves to emphasize that $\tau_i$ is a parameter on the world line rather than a coordinate, and allows us to bring many equations into a form familiar from special relativity and then benefit from known special relativistic results.

Using (11), we can define a contravariant four-velocity $v_i^\rho$ and four-acceleration $a_i^\rho$ by

$$v_i^\rho = dz_i^\rho / d\tau_i, \quad a_i^\rho = d^2 z_i^\rho / d\tau_i^2. \tag{12}$$

In the absence of a nonsingular metric, we cannot automatically define any corresponding covariant vectors. However, the vector $w_\mu$ defined in Eq. (7) can be written

$$w_\mu = g_{\mu\nu} v_i^\nu; \tag{13}$$

thus, it *might* be taken formally as the covariant vector corresponding to $v_i^\rho$. But, it should be noted that it is *not* possible to regain $v_i^\rho$ from $w_\rho$. These quantities are related by

$$v_i^\rho w_\rho = 1, \quad a_i^\rho w_\rho = 0. \tag{14}$$

In order to connect covariant and contravariant quantities, we have the following (FN 38-45)

**Lemma**: *Given a contravariant four-vector* $B^\rho = (0, \vec{B})$ *satisfying the condition*

$$B^\rho w_\rho = 0 \tag{15}$$

*and a three-velocity* $\vec{v}_i$ *(with* $v_i^m = dz_i^m / dz_i^0$*), the quantity*

$$B_{i\rho} = (\vec{B} \cdot \vec{v}_i, -\vec{B}) \tag{16}$$

*that satisfies the condition*

$$B_{i\rho} v_i^{\rho} = 0 \tag{17}$$

*[with $v_i^{\mu}$ defined as in Eq. (12)] transforms as a covariant vector under the Galilei group. Then $B^{\rho}$ is related to $B_{i\rho}$ by*

$$B^{\rho} = h^{\rho\sigma} B_{i\sigma}, \quad B_{i\rho} = k_{i\rho\sigma} B^{\sigma}, \tag{18}$$

*and the tensors $h^{\rho\sigma}$ and $k_{i\rho\sigma}$ are defined by Eq. (4) and by*

$$k_{i00} = -\vec{v}_i^{\,2}, \quad k_{i0r} = k_{ir0} = v_i^{\,r}, \quad k_{i11} = k_{i22} = k_{i33} = -1, \text{ all other } k_{i\rho\sigma} = 0 \tag{19}$$

*respectively, with*

$$v_i^{\,\rho} k_{i\rho\sigma} = 0. \tag{20}$$

We now can define the covariant four-acceleration

$$a_{i\rho} \equiv k_{i\rho\sigma} a_i^{\,\sigma} = (\vec{a}_i \cdot \vec{v}_i, -\vec{a}_i) \tag{21}$$

which satisfies

$$a_{i\rho} v_i^{\,\rho} = 0. \tag{22}$$

Although a covariant four-acceleration has been defined, no one-body covariant four-vectors can be constructed from either the position vector $\vec{r}_i$ or the velocity $\vec{v}_i$. Thus $a_{i\rho}$ is not the derivative of a four-vector.

On the other hand, since $k_{i\rho\sigma}$ is a tensor, it can be used to construct covariant vectors from *any* contravariant ones. However, it is still singular, and thus not suitable for the role of a metric tensor. It is only in the special case of contravariant vectors satisfying the condition Eq. (15) (or more generally for tensors of any rank satisfying similar conditions) that an index lowered by means of $k_{i\rho\sigma}$ can be raised to regain the original vector. This is due to the fact that

$$k_{im\sigma} h^{\sigma n} = \delta_m^{\,n}, \tag{23}$$

so that $k_{i\mu\sigma}$ and $h^{\sigma\nu}$ are reciprocals in the three-dimensional subspace formed by the $x^m$'s.

The tensor $k_{i\rho\sigma}$ is defined here (and was in FN43) explicitly by (19) and we shall find it convenient to work with this explicit three-dimensional representation in some of the following. However, in Eq. (35) below we shall give an expression for $k_{i\rho\sigma}$ in terms of other quantities (not realized in FN); this representation makes evident the tensor character of $k_{i\rho\sigma}$, as well as of Eq.(20).

Now we are ready to consider the possible invariants of the Galilei group. If we restrict ourselves to invariants that can be constructed[27] from the positions and their first, but no higher, derivatives, the vectors and tensors we have available are

$$s_{ij}^{\,\mu}(\tau_i, \tau_j) \equiv z_i^{\,\mu}(\tau_i) - z_j^{\,\mu}(\tau_j), \quad v_{ij}^{\,\mu}(\tau_i) \equiv v_i^{\,\mu}(\tau_i) - v_j^{\,\mu}(\tau_j), \quad k_{i\rho\sigma}(\tau_i), \quad i,j = 1...n \tag{24}$$

and

$$w_{\mu}, \quad g_{\mu\nu}, \quad h^{\mu\nu}, \quad \varepsilon_{\alpha\beta\gamma\delta}, \quad \varepsilon^{\alpha\beta\gamma\delta}, \quad \delta_{\mu}^{\,\nu}, \tag{25}$$

where $w_{\mu}$, $g_{\mu\nu}$, and $h^{\mu\nu}$ are tensors with numerically the same components in all coordinate systems characteristic of the Galilei group, related by Eqs. (5) and (8). The quantities $\varepsilon_{\alpha\beta\gamma\delta}$, and $\varepsilon^{\alpha\beta\gamma\delta}$ are totally antisymmetric tensor densities of weight -1 and +1 respectively, the Levi-Civita tensor densities, which have components numerically the same in all coordinate systems in any flat four-dimensional space, with $\varepsilon_{0123} = +1$, $\varepsilon^{0123} = -1$, and $\delta_{\mu}^{\,\nu}$ is the Kronecker delta, which is a mixed tensor in all spaces.[24, 25]

Clearly the only invariants of interest are those that do not vanish, and are not simply numbers *identically*, but functions of the positions and velocities. Therefore, we shall omit invariants such as $k_{i\rho\sigma}v_i^\rho v_i^\sigma$ or $v_i^\rho w_\rho$, [which equal 0 and 1 from Eqs. (20) and (14), respectively] from consideration, as well as all tensors constructed from just those listed in (25) above. Then no nontrivial *one*-particle invariants using just one event per world line can be constructed at all.

We now consider the possible two-particle invariants, for any two particles i and j, $i \ne j$. While the one-particle tensor $k_{i\rho\sigma}(\tau_i)$ is singular, the two-particle tensor

$$k_{ij\rho\sigma}(\tau_i, \tau_j) \equiv -\tfrac{1}{2}[k_{i\rho\sigma}(\tau_i) + k_{j\rho\sigma}(\tau_j)] \tag{26}$$

is *not*. Its elements are

$$k_{ij00} = \tfrac{1}{2}(\vec{v}_i^2 + \vec{v}_j^2), \quad k_{ij0m} = k_{ijm0} = -\tfrac{1}{2}(v_i^m + v_j^m),$$
$$k_{ij11} = k_{ij22} = k_{ij33} = 1, \quad \text{all other } k_{ij\rho\sigma} = 0, \tag{27}$$

and its determinant is $\tfrac{1}{4}\vec{v}_{ij}^2$, which is non-zero except for two particles having equal velocity vectors. Its inverse tensor $k_{ij}^{\rho\sigma}$ transforms as a contravariant tensor under the Galilei transformations and has the elements

$$k_{ij}^{00} = (\tfrac{1}{4}\vec{v}_{ij}^2)^{-1}, \quad k_{ij}^{0m} = k_{ij}^{m0} = \tfrac{1}{2}(v_i^m + v_j^m)/\tfrac{1}{4}\vec{v}_{ij}^2,$$
$$k_{ij}^{mm} = 1 + \frac{(v_i^m + v_j^m)^2}{\vec{v}_{ij}^2}, \quad k_{ij}^{mn} = k_{ij}^{nm} = \tfrac{1}{4}(v_i^m + v_j^m)(v_i^n + v_j^n)/\tfrac{1}{4}\vec{v}_{ij}^2, \quad n \ne m \ne 0. \tag{28}$$

As will be seen below, when used with contravariant two-particle vectors, $k_{ij\rho\sigma}$ acts like a "two-body metric tensor" for the purpose of calculating invariants. However, it cannot be used to lower indices on *one-body* contravariant vectors because such quantities would involve the variables of both particles. As noted before, there are no one-body covariant position or velocity vectors with which $k_{ij}^{\rho\sigma}$ could form invariants.

A more useful two-body "separation" four-vector $S_{ij}^\mu$ is defined (FN 37) by

$$S_{ij}^\mu \equiv s_{ij}^\rho [\delta_\rho^\mu - \tfrac{1}{2}w_\rho(v_i^\mu + v_j^\mu)] = [0, (\vec{r}_i - \vec{r}_j) - \tfrac{1}{2}(\vec{v}_i + \vec{v}_j)(t_i - t_j)], \tag{29}$$

since the three-vector part of $S_{ij}^\rho$ is Galilei invariant, while that of $s_{ij}^\rho$ is not. Both four-vectors $S_{ij}^\mu$ and $v_{ij}^\mu$ have the property like $B^\rho$ in (15) that

$$v_{ij}^\mu w_\mu = S_{ij}^\mu w_\mu = 0,$$

but not the transformations necessary for the lemma Eq. (18).

We first note that the invariant corresponding to (9) is

$$w_\mu s_{ij}^\mu = t_i - t_j \equiv t_{ij}, \tag{30}$$

which is antisymmetric in i and j. Then, the other three invariants are the obvious contractions of $k_{ij\rho\sigma}$ with $S_{ij}^\rho$ and $v_{ij}^\sigma$, viz.,

$$S_{ij}^2 \equiv k_{ij\rho\sigma} S_{ij}^\rho S_{ij}^\sigma = [\vec{r}_{ij} - \tfrac{1}{2}(\vec{v}_i + \vec{v}_j)t_{ij}]^2 = \vec{S}_{ij}^2, \tag{31}$$

$$v_{ij}^2 \equiv k_{ij\rho\sigma} v_{ij}^\rho v_{ij}^\sigma = \vec{v}_{ij}^2, \tag{32}$$

$$K_{ij} \equiv k_{ij\rho\sigma} S_{ij}^\rho v_{ij}^\sigma = k_{ij\rho\sigma} v_{ij}^\rho S_{ij}^\sigma = \vec{v}_{ij} \cdot \vec{S}_{ij}, \tag{33}$$

which are all symmetric in i and j. Only the three-dimensional forms of these particle-symmetric invariants (31)-(33) were listed in (FN37) and were offered without proof. The quantity

$$s_{ij}^2 \equiv k_{ij\rho\sigma} s_{ij}^\rho s_{ij}^\sigma = S_{ij}^2 + \tfrac{1}{4} v_{ij}^2 t_{ij}^2 \tag{34}$$

is not an independent invariant and lacks the "naturalness" of Eqs. (31)-(33). All four invariants are functions of two times and none of them depends on space coordinates only, unless the first invariant (30) vanishes—the case exclusively considered in Newtonian mechanics. Rather than treating Eqs. (30)-(33) as relating two world lines, if two points are chosen on *one* world line, these become one-particle but two-time invariants with no apparent use.

The singular tensor $k_{i\rho\sigma}$ can be written using manifestly tensorial quantities in terms of the sets (24) and (25) as

$$k_{i\rho\sigma} \equiv -\tfrac{1}{2} \varepsilon_{\rho\alpha\beta\gamma} \varepsilon_{\sigma\mu\nu\lambda} v_i^\alpha v_i^\mu h^{\beta\nu} h^{\gamma\lambda}, \tag{35}$$

yielding, therefore, for the non-singular tensor

$$k_{ij\rho\sigma} \equiv -\tfrac{1}{2}(k_{i\rho\sigma} + k_{j\rho\sigma}) = \tfrac{1}{4} \varepsilon_{\rho\alpha\beta\gamma} \varepsilon_{\sigma\mu\nu\lambda} (v_i^\alpha v_i^\mu + v_j^\alpha v_j^\mu) h^{\beta\nu} h^{\gamma\lambda}. \tag{36}$$

Alternative four-dimensional forms of the invariants (31) to (33) that are useful in calculating partial derivatives of the invariants needed in Sec. III are

$$\begin{aligned}
\vec{S}_{ij}^{\,2} &\equiv k_{ij\rho\sigma}[s_{ij}^{\,\rho} s_{ij}^{\,\sigma} - \tfrac{1}{4}(v_i^\rho v_i^\sigma + v_j^\rho v_j^\sigma)(w_\mu s_{ij}^{\,\mu})^2], \\
S_{ij}^2 &= \tfrac{1}{4} \varepsilon_{\alpha\beta\gamma\delta} \varepsilon_{\mu\nu\rho\sigma} \{(v_i^\alpha v_i^\mu + v_j^\alpha v_j^\mu) s_{ij}^\beta s_{ij}^\nu - \tfrac{1}{2} v_i^\alpha v_i^\mu v_j^\beta v_j^\nu [w_\mu s_{ij}^{\,\mu}]^2\} h^{\gamma\rho} h^{\delta\sigma}, \\
v_{ij}^2 &= -\tfrac{1}{2}(k_{i\rho\sigma} v_j^\rho v_j^\sigma + k_{j\rho\sigma} v_i^\rho v_i^\sigma), \\
\vec{v}_{ij}^{\,2} &= \tfrac{1}{2} \varepsilon_{\alpha\beta\gamma\delta} \varepsilon_{\mu\nu\rho\sigma} v_i^\alpha v_i^\mu v_j^\beta v_j^\nu h^{\gamma\rho} h^{\delta\sigma}, \\
K_{ij} &= \tfrac{1}{2}(k_{i\rho\sigma} v_j^{\,\sigma} - k_{j\rho\sigma} v_i^{\,\sigma}) S_{ij}^{\,\rho}, \\
K_{ij} &= \tfrac{1}{4} \varepsilon_{\alpha\beta\gamma\delta} \varepsilon_{\mu\nu\rho\sigma} (v_j^\beta v_j^\nu v_i^\mu - v_i^\beta v_i^\nu v_j^\mu) s_{ij}^\alpha h^{\gamma\rho} h^{\delta\sigma}.
\end{aligned} \tag{37}$$

However, there may exist other invariants. As noted by Kucharzewski[28], whenever a number of vectors is linearly dependent, the coefficients expressing the linear dependence (and functions thereof) are invariant under *linear* coordinate transformations. Usually these invariants need not be considered, as they can be expressed in terms of the invariant norms of the various vectors, if the space is metric and the norms do not vanish.[29] Since the space of the Galilei group is not metric, we must concern ourselves with this possibility. However, in general no linear dependence can exist between the three available vectors $v_i^\mu$, $v_j^\mu$, and $s_{ij}^\mu$, and thus no Galilei invariants other than (30)-(33) exist. If more vectors are available, as in the generalizations discussed below, nontrivial invariants may be provided by the coefficients.

In the above we were concerned only with invariants of the proper orthochronous Galilei group. However, obviously $\vec{v}_{ij}^{\,2}$ and $\vec{S}_{ij}^{\,2}$ are invariants of the full group, and so are the squares of $t_{ij}$ and of $K_{ij}$. Thus the full Galilei group, too, has four independent two-particle polynomial invariants.

If we wish to include accelerations (and higher derivatives), the set (25) remains unchanged, and the quantities $a_i^\mu$ and $a_j^\mu$ (and their derivatives) have to be added to the set (24). We can then proceed as before. The case of two-particle invariants including accelerations is treated in Appendix 1.

The partial derivatives required in Sec. III can be calculated using the invariants (30)-(33) or (30) and their alternative forms (37) as

$$\frac{\partial \vec{S}_{ij}^2}{\partial z_i^\mu} = 2k_{ij\mu\sigma} S_{ij}^\sigma - w_\mu k_{ij\rho\sigma}(v_i^\rho + v_j^\rho)S_{ij}^\sigma = \{-(\vec{v}_i + \vec{v}_j)\bullet\vec{S}_{ij},\ 2\vec{S}_{ij}\} = -\frac{\partial \vec{S}_{ij}^2}{\partial z_j^\mu},$$

$$\frac{\partial \vec{S}_{ij}^2}{\partial v_i^\lambda} = -\tfrac{1}{2}\varepsilon_{\lambda\beta\gamma\delta}\varepsilon_{\mu\nu\rho\sigma}s_{ij}^\beta s_{ij}^\mu v_i^\nu h^{\gamma\rho}h^{\delta\sigma} + \tfrac{1}{4}\varepsilon_{\lambda\beta\gamma\delta}\varepsilon_{\mu\nu\rho\sigma}t_{ij}^2 v_j^\beta v_j^\mu v_i^\nu h^{\gamma\rho}h^{\delta\sigma}$$

$$= \{\vec{r}_{ij}\bullet\vec{S}_{ij},\ -t_{ij}\vec{S}_{ij}\} = \frac{\partial S_{ij}^2}{\partial v_j^\lambda}$$

(38)

$$\frac{\partial \vec{v}_{ij}^2}{\partial z_i^\mu} = \frac{\partial \vec{v}_{ij}^2}{\partial z_j^\mu} = 0,\quad \frac{\partial \vec{v}_{ij}^2}{\partial v_i^\mu} = [2\vec{v}_j\bullet\vec{v}_{ji},\ 2\vec{v}_{ij}],\quad \frac{\partial \vec{v}_{ij}^2}{\partial v_j^\mu} = [2\vec{v}_i\bullet\vec{v}_{ij},\ 2\vec{v}_{ji}],$$

(39)

$$\frac{\partial K_{ij}}{\partial z_i^\mu} = -\frac{\partial K_{ij}}{\partial z_j^\mu} = \tfrac{1}{4}\varepsilon_{\alpha\beta\gamma\delta}\varepsilon_{\mu\nu\rho\sigma}(v_j^\beta v_j^\mu v_i^\alpha - v_i^\beta v_i^\mu v_j^\alpha)h^{\gamma\rho}h^{\delta\sigma} = [\tfrac{1}{2}(\vec{v}_j^2 - \vec{v}_i^2),\ \vec{v}_{ij}],$$

$$\frac{\partial K_{ij}}{\partial v_i^\lambda} = \tfrac{1}{4}\varepsilon_{\lambda\nu\rho\sigma}\varepsilon_{\alpha\beta\gamma\delta}[s_{ij}^\alpha(v_i^\beta + v_j^\beta)v_j^\nu + s_{ij}^\nu v_j^\alpha v_i^\beta]h^{\gamma\rho}h^{\delta\sigma}$$

$$= \{\tfrac{1}{2}[t_{ij}\vec{v}_j\bullet(\vec{v}_i + \vec{v}_j) + \vec{v}_{ij}\bullet\vec{r}_{ij} - 2\vec{v}_j\bullet\vec{r}_{ij}],\ \vec{r}_{ij} - \vec{v}_i t_{ij}\},$$

$$\frac{\partial K_{ij}}{\partial v_j^\mu} = \{\tfrac{1}{2}[-t_{ij}\vec{v}_i\bullet(\vec{v}_i + \vec{v}_j) + \vec{v}_{ij}\bullet\vec{r}_{ij} + 2\vec{v}_i\bullet\vec{r}_{ij}],\ -\vec{r}_{ij} + \vec{v}_j t_{ij}\}.$$

(40)

Furthermore we have obviously

$$\frac{\partial t_{ij}}{\partial z_i^\mu} = -\frac{\partial t_{ij}}{\partial z_j^\mu} = w_\mu,\qquad \frac{\partial t_{ij}}{\partial v_i^\mu} = \frac{\partial t_{ij}}{\partial v_j^\mu} = 0.$$

(41)

It should be noted that we must use the four-dimensional forms (37) to be able to evaluate derivatives with respect to $v_i^0$ or $v_j^0$ correctly; the other derivatives can be obtained from either the four- or the three-dimensional forms. Although Eqs. (38)-(41) are covariant four-vectors, they cannot be constructed via the lemma in Eqs. (15)-(20) because they have no contravariant forms that satisfy the conditions of the lemma.

### 3. Generalized Galilei-invariant mechanics

In four-dimensional notation, "Newton's second law" for particle i can be written using *contravariant* vectors (FN54)

$$m_i a_i^\rho = dP_i^\rho / d\tau_i = F_i^\rho,\quad P_i^\rho \equiv m_i v_i^\rho,\quad F_i^\rho \equiv (0, \vec{F}_i),$$

$$w_\rho dP_i^\rho / d\tau_i = w_\rho F_i^\rho = 0,$$

(42)

or equivalently using *covariant* vectors as (FN56-58)

$$m_i a_{i\rho} = dP_{i\rho} / d\tau_i = F_{i\rho},\quad F_{i\rho} = k_{i\rho\sigma}F_i^\sigma = (\vec{v}_i\bullet\vec{F}_i, -\vec{F}_i),$$

$$P_{i\rho} \equiv (T_i,\ -\vec{p}_i),\quad T_i \equiv \tfrac{1}{2}m_i\vec{v}_i^2,\quad \vec{p}_i \equiv m_i\vec{v}_i,$$

$$v_i^\rho dP_{i\rho} / d\tau_i = v_i^\rho F_{i\rho} = 0,$$

(43)

where $\tau_i$ is defined by Eq. (11), $v_i^\rho$ and $a_i^\rho$ are defined by Eq. (12) and satisfy Eq. (14), and $a_{i\rho}$ is defined by Eq. (21) and satisfies Eq. (22). While $P_i^\rho$ is a four-vector, $P_{i\rho}$ *is not*; only $dP_{i\rho}$ transforms as a covariant vector. The same physics is in the contravariant forms Eqs. (42) as in the covariant forms (43), except that in the latter, the rate of doing work is carried in the zeroth component. However, the latter is required for a Lagrangian formulation.

In the following we wish to maintain the forms (42) and (43) of the second law. Just as in the customary form of Newtonian particle dynamics, we shall assume that

$$F_i^\rho = \sum_{j \neq i} F_{ij}^\rho \big|_{g_{ij}=0}, \tag{44}$$

i.e. that we are dealing with two-body forces only. The purpose of the restriction $g_{ij} = 0$, where $g_{ij}$ is any function whatever of the two-body Galilei invariants, is to invariantly connect one or more points on the world line of particle j to a point on the world line of particle i so that the invariant parameter of i can be the ultimate independent variable in the equation of motion. One possibility is an integral over all $\tau_j$.

It might appear natural to assume that $F_{ij}^\rho$ depends only on those four-vectors and two-particle invariants of the Galilei group which are formed from the positions and velocities, i.e.,

$$F_{ij}^\rho = B_{ij} S_{ij}^\rho + C_{ij} v_{ij}^\rho \tag{45}$$

where $B_{ij}$ and $C_{ij}$ are arbitrary functions of the four invariants Eqs. (30)-(33) from Sec. 2. That the velocity dependence must involve the difference $v_{ij}^\rho$ follows from the last of Eqs. (42). In the Newtonian case $t_{ij}$ is postulated to be equal to zero, and then $\vec{S}_{ij}^2$ reduces to $\vec{r}_{ij}^2$; thus, it is possible to have two-body forces depending on positions alone, if $F_{ij}^\rho$ is assumed to depend neither on $\vec{v}_{ij}^2$ nor $K_{ij}$. If $t_{ij} \neq 0$, however, the three remaining invariants all depend on the velocities; thus, $\vec{S}_{ij}^2$ no longer has a distinguished role, and we shall therefore provisionally maintain the full dependence (45).

However, while this would be sufficient to build up a generalized *Newtonian* mechanics, it is not adequate for the construction of a generalized Lagrangian mechanics. In the customary form of Newtonian mechanics, the dependence of the forces on the simultaneous mutual separations permits the introduction of a potential energy with a similar dependence. If the forces are velocity-dependent, however, this is not possible; conversely, if the potential energy is assumed to be velocity-dependent, this leads in general to acceleration-dependent forces.[30] Since we will be interested in a generalized Lagrangian formalism, we therefore will assume instead of (45) that

$$F_{ij}^\rho = B_{ij} S_{ij}^\rho + C_{ij} v_{ij}^\rho + D_{ij} a_i^\rho + E_{ij} a_j^\rho, \tag{46}$$

where, here, $B_{ij}$ through $E_{ij}$ are functions of the two-particle invariants of the Galilei group formed from the positions, velocities, and accelerations, and can be found in Appendix 1.

In the four-dimensional formulation of Lagrangian mechanics we have (FN 78N)

$$\delta J = 0, \quad J = \int_{-\infty}^{\infty} L d\tau_i, \quad L = \tfrac{1}{2} \frac{m_i \vec{v}_i^2}{w_\mu v_i^\mu} - w_\mu v_i^\mu U_i, \tag{47}$$

so the Euler-Lagrange equations are

$$\frac{\partial L}{\partial z_i^\rho} - \frac{d}{d\tau_i} \frac{\partial L}{\partial v_i^\rho} = 0, \tag{48}$$

which lead to the covariant form (43) of the second law with

$$F_{i\rho} = \frac{\partial U_i}{\partial z_i^\rho} - \frac{d}{d\tau_i} \frac{\partial}{\partial v_i^\rho} (w_\mu v_i^\mu U_i), \tag{49}$$

taking account of (14).

In the above we have concentrated our attention on particle i alone; for several particles one can write either a joint variational principle for the particles and fields or a Fokker-type[31] variational principle for the particles alone. The first type will be of no use for our purposes, since, in general, our potentials are velocity-dependent and cannot be expected to satisfy simple partial differential equations. We shall therefore take the second type as fundamental, as indeed is its three-dimensional counterpart in the usual Lagrangian particle dynamics. The four-dimensional parameter-invariant form for the usual Newtonian-type interactions is[32] (FN 82N)

$$\delta J = 0, \quad J = \sum_i \int \tfrac{1}{2} \frac{m_i \vec{v}_i^2}{w_\mu v_i^\mu} d\tau_i - 4\pi \sum_i \sum_{<j} \iint w_\mu v_i^\mu w_\nu v_j^\nu G_{ij} d\tau_i d\tau_j, \tag{50}$$

$$G_{ij} \equiv G_{ij}(s_{ij}^\mu) = \tfrac{1}{4\pi} \delta(w_\rho s_{ij}^\rho) V_{ij}(k_{ij\rho\sigma} S_{ij}^\rho S_{ij}^\sigma);$$

in the particular case of gravitational or electrostatic interactions

$$V_{ij} = k g_i g_j (k_{ij\rho\sigma} S_{ij}^\rho S_{ij}^\sigma)^{-\tfrac{1}{2}}, \tag{51}$$

where k and $g_i$ are appropriate to the interaction. We can define adjunct potential energies

$$U_i(z_i^\mu) = \sum_{j \neq i} U_{ij}(z_i^\mu), \quad U_{ij}(z_i^\mu) \equiv -4\pi \int_{-\infty}^{\infty} w_\nu v_j^\nu G_{ij}(s_{ij}^\mu) d\tau_j; \tag{52}$$

then the Euler-Lagrange equations for the i-th particle take the form (48) with (49) as before.

For arbitrary coordinates (not necessarily corresponding to the location of a particle), $G_{ij}$ could be considered as a Green's function for some partial differential equation satisfied by the potential energy defined by

$$U(x^\mu) = -4\pi \int \rho(x'^\mu) G(x'^\mu - x^\mu) d^4 x', \tag{53}$$

where $d^4 x'$ is the four-dimensional volume element and the density is

$$\rho = \sum_{i=1}^n g_i \int_{-\infty}^{\infty} \delta^4[x'^\rho - z_i^\rho(\tau_i)] d\tau_i \tag{54}$$

for point particles. In the following, form (50) will be generalized. To simplify the algebra, we shall replace the parameter-invariant variational principle (50) by the parameter-dependent form

$$J = \sum_i \int_{-\infty}^{\infty} \tfrac{1}{2} \frac{m_i \vec{v}_i^2}{w_\mu v_i^\mu} d\tau_i - 4\pi \sum_{i<} \sum_j \iint G_{ij}(s_{ij}^\alpha, v_i^\alpha, v_j^\alpha) d\tau_i d\tau_j, \tag{55}$$

where now the variation has to be performed subject to the subsidiary condition (14) on $v_i^\mu$, or equivalently

$$w_\mu \delta v_i^\mu = 0. \tag{56}$$

Therefore, in performing the variation of (55), we have to add to $\delta J$ the expression

$$\sum_i \int_{-\infty}^{\infty} d\tau_i Q_i w_\mu \delta v_i^\mu, \tag{57}$$

where the $Q_i$ are Lagrange multipliers which have to be determined from the resulting equations. Before giving the results of this calculation, we shall first introduce a generalization of $G_{ij}$ in (50). The double integral in (55) is clearly Galilei invariant[33], since so are both functions in $G_{ij}$ of (50). It will remain Galilei invariant, if, instead of the special function (50), we introduce an arbitrary function

$$G_{ij} = G_{ij}(t_{ij}, S_{ij}^2, v_{ij}^2, K_{ij}) \tag{58}$$

of the four invariants found in Sec. 2. Therefore we shall base our generalized Lagrangian Galilei-invariant dynamics on the variational principle (55) with (57) and (58). Because of the form (30) of $t_{ij}$, $G_{ij}$ no longer is necessarily symmetric in i and j. For n particles, the interactions are characterized by $\frac{1}{2}n(n-1)$ possibly distinct functions $G_{ij}$.

The appearance of more complicated interactions than those described by (50) is the principle reason for the use of a four-dimensional formalism in this paper; a three-dimensional variational principle is not well suited for the description of interactions which do not depend on the same time for all particles. This is in addition to the general reason of wishing to stress similarities with special relativistic dynamics. We shall continue to refer to any function $G_{ij}$ appearing in (55) as a Green's function, whether or not it can serve as such a function for a partial differential equation. The adjunct potential energy $U_i$ defined in (52) is the potential energy (53) evaluated at $z_i^\mu$, with the contribution of the i-th particle omitted. For our new Green's functions (58) we can still define adjunct one-body potential energies

$$U_i(z_i^\mu, v_i^\mu) = \sum_{j<i}(-4\pi)\int G_{ji}d\tau_j + \sum_{j>i}(-4\pi)\int G_{ij}d\tau_j, \qquad (59)$$

where two sums must be introduced because of the possible lack of particle-symmetry of $G_{ij}$.

Unlike the potential energies (52), the expressions (59) depend on the velocity of the i-th particle [at the position where (59) is evaluated], except for the somewhat trivial case of $G_{ij}$ being a function of $t_{ij}$ alone, and the Newtonian case $t_{ij} = 0$. Therefore, excepting these cases, it is no longer possible to define a potential energy (53) at any arbitrary point.

Now we carry out the variation[34] of Eq. (55), adding (57), resulting in the equations of motion

$$\frac{dP_{i\rho}}{d\tau_i} = -\frac{\partial U_i}{\partial z_i^\rho} + \frac{d}{d\tau_i}\left\{\frac{\partial U_i}{\partial v_i^\rho} + w_\rho\left[U_i - v_i^\sigma\frac{\partial U_i}{\partial v_i^\sigma}\right]\right\} \equiv F_{i\rho}, \qquad (60)$$

ignoring an irrelevant constant of integration. More explicitly, the forces in (60), with $G_{ij}$ given by (58), have the form

$$\begin{aligned}
F_{i\rho} = &-4\pi\sum_{j>i}\int_{-\infty}^{\infty}d\tau_j\Bigg\{(-1)\left(\frac{\partial G_{ij}}{\partial t_{ij}}\frac{\partial t_{ij}}{\partial z_i^\rho} + \frac{\partial G_{ij}}{\partial \vec{S}_{ij}^2}\frac{\partial \vec{S}_{ij}^2}{\partial z_i^\rho} + \frac{\partial G_{ij}}{\partial \vec{v}_{ij}^2}\frac{\partial \vec{v}_{ij}^2}{\partial z_i^\rho} + \frac{\partial G_{ij}}{\partial K_{ij}}\frac{\partial K_{ij}}{\partial z_i^\rho}\right) \\
&+\frac{d}{d\tau_i}\left(\frac{\partial G_{ij}}{\partial t_{ij}}\frac{\partial t_{ij}}{\partial v_i^\rho} + \frac{\partial G_{ij}}{\partial \vec{S}_{ij}^2}\frac{\partial \vec{S}_{ij}^2}{\partial v_i^\rho} + \frac{\partial G_{ij}}{\partial \vec{v}_{ij}^2}\frac{\partial \vec{v}_{ij}^2}{\partial v_i^\rho} + \frac{\partial G_{ij}}{\partial K_{ij}}\frac{\partial K_{ij}}{\partial v_i^\rho}\right) \\
&+w_\rho\frac{d}{d\tau_i}\left[G_{ij} - v_i^\sigma\left(\frac{\partial G_{ij}}{\partial t_{ij}}\frac{\partial t_{ij}}{\partial v_i^\sigma} + \frac{\partial G_{ij}}{\partial \vec{S}_{ij}^2}\frac{\partial \vec{S}_{ij}^2}{\partial v_i^\sigma} + \frac{\partial G_{ij}}{\partial \vec{v}_{ij}^2}\frac{\partial \vec{v}_{ij}^2}{\partial v_i^\sigma} + \frac{\partial G_{ij}}{\partial K_{ij}}\frac{\partial K_{ij}}{\partial v_i^\sigma}\right)\right]\Bigg\} \\
&-4\pi\sum_{j<i}\int_{-\infty}^{\infty}d\tau_j\Bigg\{(-1)\left(\frac{\partial G_{ji}}{\partial t_{ji}}\frac{\partial t_{ji}}{\partial z_i^\rho} + \frac{\partial G_{ji}}{\partial \vec{S}_{ji}^2}\frac{\partial \vec{S}_{ji}^2}{\partial z_i^\rho} + \frac{\partial G_{ji}}{\partial \vec{v}_{ji}^2}\frac{\partial \vec{v}_{ji}^2}{\partial z_i^\rho} + \frac{\partial G_{ji}}{\partial K_{ji}}\frac{\partial K_{ji}}{\partial z_i^\rho}\right) \\
&+\frac{d}{d\tau_i}\left(\frac{\partial G_{ji}}{\partial t_{ji}}\frac{\partial t_{ji}}{\partial v_i^\rho} + \frac{\partial G_{ji}}{\partial \vec{S}_{ji}^2}\frac{\partial \vec{S}_{ji}^2}{\partial v_i^\rho} + \frac{\partial G_{ji}}{\partial \vec{v}_{ji}^2}\frac{\partial \vec{v}_{ji}^2}{\partial v_i^\rho} + \frac{\partial G_{ji}}{\partial K_{ji}}\frac{\partial K_{ji}}{\partial v_i^\rho}\right) \\
&+w_\rho\frac{d}{d\tau_i}\left[G_{ji} - v_i^\sigma\left(\frac{\partial G_{ji}}{\partial t_{ji}}\frac{\partial t_{ji}}{\partial v_i^\sigma} + \frac{\partial G_{ji}}{\partial \vec{S}_{ji}^2}\frac{\partial \vec{S}_{ji}^2}{\partial v_i^\sigma} + \frac{\partial G_{ji}}{\partial \vec{v}_{ji}^2}\frac{\partial \vec{v}_{ji}^2}{\partial v_i^\sigma} + \frac{\partial G_{ji}}{\partial K_{ji}}\frac{\partial K_{ji}}{\partial v_i^\sigma}\right)\right]\Bigg\},
\end{aligned} \qquad (61)$$

where the derivatives of the invariants are given by Eqs. (38)-(41). Clearly, in general, these forces (61) [or Eq. (44) with (46)] are acceleration dependent and do not obey even a generalized form of Newton's third law of motion. Furthermore, the equivalence of the instantaneous forms of Newtonian and Lagrangian mechanics does not carry through to generalized Galilei-invariant classical mechanics as a comparison of Eqs. (60) with (61) versus (42) with (44) and (46) makes clear.

In Eq. (58) the form of the Green's function has been left completely arbitrary. However, the appearance of two factors in the Newtonian Green's function (50) has a fundamental physical significance, obscured by our familiarity with Newtonian mechanics. An interaction is characterized by two laws, one expressing the dependence of the force on the particle variables of the particles under consideration and a second law specifying which points on the world line must be used in evaluating the particle variables.[35] In Newtonian mechanics the first type of law is the usual force law expressing the dependence on the mutual particle separations; the second type states that the positions are to be evaluated at the same time. In the interactions encountered in electrodynamics the second type of law relates a point on one world line with the retarded (or advanced) point on the other,[31] and conversely; in mesodynamics, a point on one world line is related with all past (or future) points with time-like separation on the other, and conversely;[36] for other relativistic interactions, space-like separations can also be considered.[35-39]

Thus, if, in $G_{ij}$ of Eq. (58) we leave a single function defined for all values of $t_{ij}$, this implies due to the form (61) that the force exerted on the i-th particle at $z_i^\mu(\tau_i)$ by the j-th one at $z_j^\mu(\tau_j)$ depends on the variables of the j-th particle at *all* $\tau_j$, and conversely. While this is certainly a possible choice of the second type of law considered above, it seems important to study the possibility of more restrictive laws. In Sec. 4 and 5, we shall therefore consider the case

$$4\pi G_{ij}(t_{ij}, \vec{S}_{ij}^2, \vec{v}_{ij}^2, K_{ij}) = V_{ij}(t_{ij}, \vec{S}_{ij}^2, \vec{v}_{ij}^2, K_{ij}) f_{ij}(t_{ij}), \tag{62}$$

where $V_{ij}$ is defined for all values of $t_{ij}$, but $f_{ij}$ restricts the range of allowed values of $t_{ij}$. We shall be particularly interested in the choice

$$4\pi G_{ij} = V_{ij}(t_{ij}, S_{ij}^2, v_{ij}^2, K_{ij}) \delta(t_{ij} - h_{ij}), \tag{63}$$

$h_{ij}$ a constant. Special cases of $f_{ij}$ to be considered are the time-asymmetric case

$$4\pi G_{ij} = V_{ij}(t_{ij}, \vec{S}_{ij}^2, \vec{v}_{ij}^2, K_{ij}) \delta(t_{ij} - T_{ij}), \tag{64}$$

and the time-symmetric case

$$4\pi G_{ij} = V_{ij}(t_{ij}, \vec{S}_{ij}^2, \vec{v}_{ij}^2, K_{ij}) \delta(t_{ij}^2 - T_{ij}^2), \tag{65}$$

where $T_{ij}$ is a constant. The use of (64) in $U_i$, defined in (59), leads (for j>i) to a dependence of the force on i due to j at an earlier (later) time if $T_{ij}$ is positive (negative), in analogy to the retarded (advanced) potential of electro-[31] and meso-dynamics[36], and conversely for j<i. Using (65) leads to a dependence on both earlier and later times, in analogy to the time-symmetric potential. The use of (64) in the variational principle (55) yields a theory which is not time symmetric, in analogy to the principle considered by Fokker in the first paper of Ref. 31 (one particle exerting retarded effects on the second, the second advanced effects on the first), while the use of (65) yields a theory which is time-symmetric, in analogy to the usual Fokker principle introduced in the second paper of Ref. 31.

It is not necessary to choose $f_{ij}$ to be a function of $t_{ij}$ alone, as in (64) and (65). However, because of the form of the other invariants, in general, any expression involving them will have an indeterminate number of roots (the number depending on the form of the world lines) for $t_{ij}$. Apart from results that can be obtained independent of the form of the Green's function, we shall therefore concentrate our attention on special cases where $f_{ij} = f_{ij}(t_{ij})$ alone.

**4. Conservation Laws**

The equations of motion (60) with (61) are the consequences of a variational principle (55) with (57) and (58), where the integral is invariant up to a divergence under the 10-parameter group (1) with (3). Therefore Noether's theorem[2-4, 14, 15] assures us of the existence of 10 conservation laws. From any of these approaches,[14, 15] we find

$$P_\mu(\tau_i^*,...,\tau_n^*) = \sum_i \left[ P_{i\mu}(\tau_i^*) - \frac{\partial U_i}{\partial v_i^\mu} - w_\mu \left( U_i - v_i^\rho \frac{\partial U_i}{\partial v_i^\rho} \right) \right]$$
$$+ \sum_{i<} \sum_j 4\pi \left[ \int_{\tau_i^*}^\infty \int_{-\infty}^{\tau_j^*} - \int_{-\infty}^{\tau_i^*} \int_{\tau_j^*}^\infty \right] d\tau_i d\tau_j \frac{\partial G_{ij}}{\partial s_{ij}^\mu},$$
(66)

and

$$L^{\mu\nu}(\tau_i^*,...,\tau_n^*) = \sum_i z_i^{[\mu}(\tau_i^*) \left\{ P_i^{\nu]}(\tau_i^*) - h^{\nu]\rho} \left[ \frac{\partial U_i}{\partial v_i^\rho} + w_\rho \left( U_i - v_i^\sigma \frac{\partial U_i}{\partial v_i^\sigma} \right) \right] \right\}$$
$$+ \tfrac{1}{2} \sum_{i<} \sum_j 4\pi \left[ \int_{\tau_i^*}^\infty \int_{-\infty}^{\tau_j^*} - \int_{-\infty}^{\tau_i^*} \int_{\tau_j^*}^\infty \right] d\tau_i d\tau_j \left\{ \left[ z_i^{[\mu} h^{\nu]\rho} \frac{\partial G_{ij}}{\partial z_i^\rho} + v_i^{[\mu} h^{\nu]\rho} \frac{\partial G_{ij}}{\partial v_i^\rho} \right] \right.$$
$$\left. - \left[ z_j^{[\mu} h^{\nu]\rho} \frac{\partial G_{ij}}{\partial z_j^\rho} + v_j^{[\mu} h^{\nu]\rho} \frac{\partial G_{ij}}{\partial v_j^\rho} \right] \right\},$$
(67)

where $G_{ij}$ is defined by (58), $U_i$ by (59), and Eqs. (66) and (67) satisfy

$$\frac{\partial P_\mu}{\partial \tau_k^*} = 0, \qquad \frac{\partial L^{\mu\nu}}{\partial \tau_k^*} = 0,$$
(68)

for any of the n values of k by utilizing the equations of motion (60). The form of the conservation laws (66) and (67) is the same as that for Poincaré-invariant theories involving integrals over the past and future world lines, but is fundamentally different from that familiar from Newtonian mechanics.

We also recall that the $P_{i\mu}$ are not covariant vectors, only the $\Delta P_{i\mu}$. Thus the vanishing of the *change* in (66) is a true covariant law, while (66) itself (which was deduced from it) is not.

All these conservation laws involve integrations over the world lines, and the Lagrangian expressions are valid for any $G_{ij}$ of the form (58). On the other hand, we know that at least in the case of Newtonian mechanics with $G_{ij}$ given by (50), the conservation laws can be put into a form not involving integrations over the world lines, i.e. not requiring a knowledge of the motion, but only of the positions and velocities at n points. The problem of finding the most general form of (58) which allows such a formulation is not attempted here. Instead, we will give only an example (including Newtonian mechanics as a special case) of such a $G_{ij}$, namely the retarded-advanced form (64) supplemented by additional conditions on the $T_{ij}$ and $\tau_i^*$. Substituting (64) into (66) we obtain

$$P_\mu(\tau_i^*,...,\tau_n^*) = \sum_i \left\{ P_{i\mu}(\tau_i^*) - w_\mu \left[ U_i - v_i^\rho \frac{\partial U_i}{\partial v_i^\rho} \right] \right\}$$
$$+ \sum_{i<j} \int_{-\infty}^\infty dt_j \delta(t_i^* - t_j - T_{ij}) \frac{\partial V_{ij}}{\partial v_i^\mu} + \sum_{i<j} \int_{-\infty}^\infty dt_i \delta(t_i - t_j^* - T_{ij}) \frac{\partial V_{ij}}{\partial v_j^\mu} \quad (69)$$
$$+ \sum_{i<j} \left[ \int_{\tau_i^*}^\infty \int_{-\infty}^{\tau_j^*} - \int_{-\infty}^{\tau_i^*} \int_{\tau_j^*}^\infty \right] d\tau_i d\tau_j \left[ \delta(t_{ij} - T_{ij}) \frac{\partial V_{ij}}{\partial s_{ij}^\mu} + w_\mu V_{ij} \frac{d\delta(t_{ij} - T_{ij})}{d(t_{ij} - T_{ij})} \right].$$

From Eq. (11), $dt_i$ equals $d\tau_i$ for all i. Choosing the arbitrary constants[26] $C_i$ each to be zero[40] results in

$$\tau_i = t_i, \quad i = 1,...,n. \quad (70)$$

Using (70) we can integrate the last term of (69) by parts, say, over $t_i$, to obtain

$$-\sum_{i<j} \int_{-\infty}^\infty dt_j w_\mu V_{ij}[t_i^* - t_j, S_{ij}^2(t_i^*,t_j), v_{ij}^2(t_i^*,t_j), K_{ij}(t_i^*,t_j)] \delta(t_j - t_i^* + T_{ij})$$
$$+ \sum_{i<j} \left[ \int_{t_i^*}^\infty \int_{-\infty}^{t_j^*} - \int_{-\infty}^{t_i^*} \int_{t_j^*}^\infty \right] dt_i dt_j \delta(t_{ij} - T_{ij}) \left[ \partial V_{ij}/\partial s_{ij}^\mu - w_\mu dV_{ij}/dt_i \right], \quad (71)$$

where the terms in the single integral that were evaluated at plus and minus infinity have vanished by the properties of the delta function. For brevity, the argument of the double integral in (71) will be designated

$$\partial V_{ij}/\partial s_{ij}^\mu - w_\mu dV_{ij}/dt_i \equiv B_{ij\mu} \equiv B_{ij\mu}[z_i^\alpha(t_i), z_j^\alpha(t_j), v_i^\alpha(t_i), v_j^\alpha(t_j), a_i^\alpha(t_i)]. \quad (72)$$

Inserting (71) with (70) and (72) in (69) and adding zero to the double-integral term in the form

$$\sum_{i<j} \left[ \int_{-\infty}^{t_i^*} \int_{-\infty}^{t_j^*} - \int_{-\infty}^{t_i^*} \int_{-\infty}^{t_j^*} \right] dt_i dt_j \delta(t_{ij} - T_{ij}) B_{ij\mu}$$

yields

$$P_\mu(t_i^*,...,t_n^*) = \sum_i \left\{ P_{i\mu}(t_i^*) - w_\mu \left[ U_i - v_i^\rho \frac{\partial U_i}{\partial v_i^\rho} \right] \right\} + \sum_{i<j} \int_{-\infty}^\infty dt_i \delta(t_i - t_j^* - T_{ij}) \left( \frac{\partial V_{ij}}{\partial v_j^\mu} \right)$$
$$+ \sum_{i<j} \int_{-\infty}^\infty dt_j \delta(t_i^* - t_j - T_{ij}) \left( \frac{\partial V_{ij}}{\partial v_i^\mu} - w_\mu V_{ij} \right) \quad (73)$$
$$+ \sum_{i<j} \left[ \int_{-\infty}^\infty \int_{-\infty}^{t_j^*} - \int_{-\infty}^{t_i^*} \int_{-\infty}^\infty \right] dt_i dt_j \delta(t_{ij} - T_{ij}) B_{ij\mu}.$$

Proceeding similarly with (67) and noting (8), we get directly

$$L^{\mu\nu}(t_i^*,...,t_n^*) = \sum_i z_i^{[\mu}(t_i^*) \left[ P_i^{\nu]}(t_i^*) - h^{\nu]\rho} \frac{\partial U_i}{\partial v_i^\rho} \right] + \tfrac{1}{2} \sum_{i<j} \left[ \int_{-\infty}^\infty \int_{-\infty}^{t_j^*} - \int_{-\infty}^{t_i^*} \int_{-\infty}^\infty \right] dt_i dt_j \delta(t_{ij} - T_{ij}) C_{ij}^{\mu\nu},$$
$$C_{ij}^{\mu\nu} \equiv \left[ z_i^{[\mu} h^{\nu]\rho} \frac{\partial V_{ij}}{\partial z_i^\rho} + v_i^{[\mu} h^{\nu]\rho} \frac{\partial V_{ij}}{\partial v_i^\rho} \right] - \left[ z_j^{[\mu} h^{\nu]\rho} \frac{\partial V_{ij}}{\partial z_j^\rho} + v_j^{[\mu} h^{\nu]\rho} \frac{\partial V_{ij}}{\partial v_j^\rho} \right]. \quad (74)$$

From the form (64) of the retarded (advanced) Green's function it follows that if, e.g., particle 1 at $t_1$ interacts with particle 2, which in turn interacts with particle 3, the times of interaction of these particles are $t_2 = t_1 - T_{12}$ and $t_3 = t_2 - T_{23}$, respectively. On the other hand, particle 1 also interacts with 3 directly, and its action is felt at a time $t_1 - T_{13}$ which could be *different* from the time obtained before unless $T_{13} = T_{12} + T_{23}$. Similarly, there will be other chains of interaction among the different particles, which will involve various different times for each particle. If we wish each particle to be involved at only one event per world line for all these chains, we must require that

$$T_{ij} = T_{ik} + T_{kj}, \quad i,j,k = 1,...,n. \tag{75}$$

This follows when the $T_{ij}$ are chosen to be differences

$$T_{ij} \equiv T_i - T_j, \quad i,j = 1,...,n, \tag{76}$$

where the $T_i$ are n arbitrary constants.

We now assume that the interactions $G_{ij}$ of form (64) have values of $T_{ij}$ that satisfy (75) ensured by (76). Furthermore, we choose to evaluate the conservation laws (73) and (74) at times $t_i^*$ such that

$$t_i^* - T_i = t_j^* - T_j, \quad i,j = 1,...,n. \tag{77}$$

In the conservation expressions (73) and (74), all integrals involve delta functions and the choice (77) assures us that the remaining double integrals will vanish and the expressions will involve only n correlated times, as now will be shown.

Performing the integrals from $-\infty$ to $\infty$ in the double integrals of (73) using (72) gives

$$\sum_{i<}\sum_{j}\int_{-\infty}^{t_j^*} dt_j B_{ij\mu}\left[z_i^\alpha(t_j+T_i-T_j), z_j^\alpha(t_j), v_i^\alpha(t_j+T_i-T_j), v_j^\alpha(t_j), a_i^\alpha(t_j+T_i-T_j)\right]$$

$$-\sum_{i<}\sum_{j}\int_{-\infty}^{t_i^*} dt_i B_{ij\mu}\left[z_i^\alpha(t_i), z_j^\alpha(t_i-T_i+T_j), v_i^\alpha(t_i), v_j^\alpha(t_i-T_i+T_j), a_i^\alpha(t_i)\right].$$

In the second expression, a change of integration variable [similar to (77)]

$$t_i = t_j + T_i - T_j \tag{78}$$

results in

$$\sum_{i<}\sum_{j}\int_{-\infty}^{t_j^*} dt_j B_{ij\mu}\left[z_i^\alpha(t_j+T_i-T_j), z_j^\alpha(t_j), v_i^\alpha(t_j+T_i-T_j), v_j^\alpha(t_j), a_i^\alpha(t_j+T_i-T_j)\right]$$

$$-\sum_{i<}\sum_{j}\int_{-\infty}^{t_i^*-T_{ij}} dt_j B_{ij\mu}\left[z_i^\alpha(t_j+T_i-T_j), z_j^\alpha(t_j), v_i^\alpha(t_j+T_i-T_j), v_j^\alpha(t_j), a_i^\alpha(t_j+T_i-T_j)\right],$$

which vanishes upon choosing the upper limit of the second integral to satisfy (77). Eq. (74) for $L^{\mu\nu}$ is handled similarly.

Thus, the difference of double integrals in the conservation laws (73) and (74) do not contribute, and these conserved quantities reduce to

$$P_\mu(t_1^*,...,t_n^*) = \sum_i P_{i\mu}(t_i^*) + w_\mu \sum_{i<}\sum_j \left[V_{ij} - \left(v_i^\rho \frac{\partial V_{ij}}{\partial v_i^\rho} + v_j^\rho \frac{\partial V_{ij}}{\partial v_j^\rho}\right)\right]\Big|^{t_i^*}$$

$$+ \sum_{i<}\sum_j \left(\frac{\partial V_{ij}}{\partial v_i^\mu} + \frac{\partial V_{ij}}{\partial v_j^\mu}\right)\Big|^{t_i^*}, \tag{79}$$

and

$$L^{\mu\nu}(t_1^*,...,t_n^*) = \sum_i z_i^{[\mu}(t_i^*) P_i^{\nu]}(t_i^*) + \sum_{i<}\sum_j \left\{z_i^{[\mu}(t_i^*) h^{\nu]\rho}\frac{\partial V_{ij}}{\partial v_i^\rho} + z_j^{[\mu}(t_j^*) h^{\nu]\rho}\frac{\partial V_{ij}}{\partial v_j^\rho}\right\}\Big|^{t_i^*}, \tag{80}$$

where the times of evaluation $t_i^*$ are restricted by (77) and the potential $V_{ij}|^{t_i^*}$ is given by

$$V_{ij}|^{t_i^*} \equiv V_{ij}\left[T_i - T_j, S_{ij}^2(t_i^*, t_i^* - T_i + T_j), v_{ij}^2(t_i^*, t_i^* - T_i + T_j), K_{ij}(t_i^*, t_i^* - T_i + T_j)\right]. \tag{81}$$

It should be noted that each of the conserved quantities (79) and (80) with (81) *really* depend only on one time, since once the n values of $T_i$ and, say, $t_1^*$ are chosen, all other times are determined by Eq. (77). Translating that one point along its world line causes all the other correlated evaluation points to move in lock step with it.

The case $\mu = 0$ in Eq. (79) gives the total energy

$$E(t_1^*,...,t_n^*) = P_0(t_1^*,...,t_n^*) = \sum_i \tfrac{1}{2}m_i \vec{v}_i^2(t_i^*) + \sum_{i<} \sum_j V_{ij}|^{t_i^*}$$
$$- \sum_{i<} \sum_j \left[\vec{v}_i(t_i^*) \cdot \frac{\partial V_{ij}}{\partial \vec{v}_i} + \vec{v}_j(t_j^*) \cdot \frac{\partial V_{ij}}{\partial \vec{v}_j}\right]|^{t_i^*}, \tag{82}$$

while choosing $\mu = m = 1,2,3$ in Eq. (79) gives the negative [recalling Eq. (43)] of the translational momentum

$$\vec{P}(t_1^*,...,t_n^*) = \sum_i \vec{p}_i(t_i^*) - \sum_{i<} \sum_j \left(\frac{\partial V_{ij}}{\partial \vec{v}_i} + \frac{\partial V_{ij}}{\partial \vec{v}_j}\right)|^{t_i^*}. \tag{83}$$

Choosing $\mu = 0$ and $\nu = n = 1,2,3$ in Eq. (80) yields the negative of the center-of-mass conserved quantity

$$\vec{G}(t_1^*,...,t_n^*) \equiv -L^{0n}(t_1^*,...,t_n^*) = \sum_i \left[m_i \vec{r}_i(t_i^*) - t_i^* \vec{p}_i(t_i^*)\right] + \sum_{i<} \sum_j \left[t_i^* \frac{\partial V_{ij}}{\partial \vec{v}_i} + t_j^* \frac{\partial V_{ij}}{\partial \vec{v}_j}\right]|^{t_i^*}, \tag{84}$$

while the choice $\mu = m$ and $\nu = n \neq m$ results in the angular momentum

$$\vec{L}(t_1^*,...,t_n^*) = \sum_i \vec{r}_i(t_i^*) \times \vec{p}_i(t_i^*) - \sum_{i<} \sum_j \left\{\vec{r}_i(t_i^*) \times \frac{\partial V_{ij}}{\partial \vec{v}_i} + \vec{r}_j(t_j^*) \times \frac{\partial V_{ij}}{\partial \vec{v}_j}\right\}|^{t_i^*}. \tag{85}$$

The partial derivatives appearing in Eqs. (82)-(85) must be evaluated with the help of Eqs. (38)-(41) giving

$$\frac{\partial V_{ij}}{\partial \vec{v}_i}|^{t_i^*} = \frac{\partial V_{ij}}{\partial S_{ij}^2}\left[-(T_i - T_j)\vec{r}_{ij} + \tfrac{1}{2}(\vec{v}_i + \vec{v}_j)(T_i - T_j)^2\right]$$
$$+ \frac{\partial V_{ij}}{\partial v_{ij}^2}\left[2\vec{v}_{ij}\right] + \frac{\partial V_{ij}}{\partial K_{ij}}\left[\vec{r}_{ij} - (\vec{v}_i)(T_i - T_j)\right]|^{t_i^*},$$

$$\frac{\partial V_{ij}}{\partial \vec{v}_j}|^{t_i^*} = \frac{\partial V_{ij}}{\partial S_{ij}^2}\left[-(T_i - T_j)\vec{r}_{ij} + \tfrac{1}{2}(\vec{v}_i + \vec{v}_j)(T_i - T_j)^2\right]$$
$$+ \frac{\partial V_{ij}}{\partial v_{ij}^2}\left[2\vec{v}_{ji}\right] + \frac{\partial V_{ij}}{\partial K_{ij}}\left[-\vec{r}_{ij} + (\vec{v}_j)(T_i - T_j)\right]|^{t_i^*}. \tag{86}$$

In Eqs. (86), the expression $\vec{r}_{ij}$, e.g., means $\vec{r}_i(t_i^*) - \vec{r}_j(t_j^*) = \vec{r}_i(t_i^*) - \vec{r}_j(t_i^* - T_i + T_j)$, having used Eq. (77); the full dependencies in Eqs. (86) have been suppressed for the sake of brevity. Inserting Eqs. (86) in Eqs. (82)-(85) yields

$$E = E(t_1^*,...,t_n^*) = \sum_i \tfrac{1}{2} m_i \vec{v}_i^{\,2}(t_i^*) + \sum_{i<} \sum_j V_{ij} \big|^{t_i^*}$$

$$- \sum_{i<} \sum_j \left\{ \frac{\partial V_{ij}}{\partial S_{ij}^2} \Big[ -(\vec{v}_i + \vec{v}_j)\bullet\vec{r}_{ij}(T_i - T_j) + \tfrac{1}{2}(\vec{v}_i + \vec{v}_j)^2 (T_i - T_j)^2 \Big] \big|^{t_i^*} \right\}$$

$$- \sum_{i<} \sum_j \left\{ + \frac{\partial V_{ij}}{\partial v_{ij}^2} 2\vec{v}_{ij}^{\,2} + \frac{\partial V_{ij}}{\partial K_{ij}} \Big[ \vec{v}_{ij}\bullet\vec{r}_{ij} - (\vec{v}_i^{\,2} - \vec{v}_j^{\,2})(T_i - T_j) \Big] \right\} \big|^{t_i^*}$$

$$\vec{P}(t_1^*,...,t_n^*) = \sum_i \vec{p}_i(t_i^*) - \sum_{i<} \sum_j \left\{ \frac{\partial V_{ij}}{\partial S_{ij}^2} \Big[ -2\vec{r}_{ij}(T_i - T_j) + (\vec{v}_i + \vec{v}_j)(T_i - T_j)^2 \Big] \right\} \big|^{t_i^*}$$

$$+ \sum_{i<} \sum_j \frac{\partial V_{ij}}{\partial K_{ij}} \Big[ (\vec{v}_{ij})(T_i - T_j) \Big] \big|^{t_i^*},$$

$$\vec{G}(t_1^*, t_2^*,...,t_n^*) = \sum_i \Big[ m_i \vec{r}_i(t_i^*) - t_i^* \vec{p}_i(t_i^*) \Big] + \sum_{i<} \sum_j \frac{\partial V_{ij}}{\partial v_{ij}^2} 2(\vec{v}_{ij})(T_i - T_j) \big|^{t_i^*},$$

$$+ \sum_{i<} \sum_j \left\{ \frac{\partial V_{ij}}{\partial S_{ij}^2} \Big[ -\vec{r}_{ij} + \tfrac{1}{2}(\vec{v}_i + \vec{v}_j)(T_i - T_j) \Big](T_i - T_j)(t_i^* + t_j^*) + \frac{\partial V_{ij}}{\partial K_{ij}} \Big[ \vec{r}_{ij} - \vec{v}_i t_i^* + \vec{v}_j t_j^* \Big](T_i - T_j) \right\} \big|^{t_i^*},$$

$$\vec{L}(t_1^*, t_2^*,...,t_n^*) = \sum_i \vec{r}_i(t_i^*) \times \vec{p}_i(t_i^*) - \sum_{i<} \sum_j \frac{\partial V_{ij}}{\partial v_{ij}^2} 2\vec{r}_{ij} \times \vec{v}_{ij} \quad (87)$$

$$- \sum_{i<} \sum_j \left\{ \frac{\partial V_{ij}}{\partial S_{ij}^2} \Big[ 2\vec{r}_i \times \vec{r}_j + \tfrac{1}{2}(\vec{r}_i + \vec{r}_j) \times (\vec{v}_i + \vec{v}_j) T_{ij} \Big](T_i - T_j) + \frac{\partial V_{ij}}{\partial K_{ij}}(-\vec{r}_i \times \vec{v}_i + \vec{r}_j \times \vec{v}_j)(T_i - T_j) \right\} \big|^{t_i^*}.$$

In the special case of Newtonian Mechanics, all times are equal ($T_i = T_j = 0$) and therefore these expressions (87) reduce to

$$E(t) = \sum_i \tfrac{1}{2} m_i \vec{v}_i^{\,2}(t) + \sum_{i<} \sum_j \left[ V_{ij} - \frac{\partial V_{ij}}{\partial v_{ij}^2} 2v_{ij}^2 - \frac{\partial V_{ij}}{\partial K_{ij}} \vec{v}_{ij} \bullet \vec{r}_{ij} \right],$$

$$\vec{P}(t) = \sum_i \vec{p}_i(t),$$

$$\vec{G}(t) = \sum_i \Big[ m_i \vec{r}_i(t) - t\vec{p}_i(t) \Big], \quad (88)$$

$$\vec{L}(t) = \sum_i \vec{r}_i(t) \times \vec{p}_i(t) - \sum_{i<} \sum_j \frac{\partial V_{ij}}{\partial v_{ij}^2} 2\vec{r}_{ij} \times \vec{v}_{ij},$$

where now

$$V_{ij} \equiv V_{ij}\Big[ 0, \vec{r}_{ij}^{\,2}(t), \vec{v}_{ij}^{\,2}(t), \vec{v}_{ij}(t) \bullet \vec{r}_{ij}(t) \Big].$$

In addition, if the potential energies do not depend on the velocities, the first and last of Eqs. (88) further reduce to

$$E(t) = \sum_i \tfrac{1}{2} m_i \vec{v}_i^2 + \sum_{i<}\sum_j V_{ij}(\vec{r}_{ij}^2),$$

$$\vec{L}(t) = \sum_i \vec{r}_i(t) \times \vec{p}_i(t),$$
(89)

where the last term of E(t) is the usual Newtonian potential energy. The extra terms in the first and last of Eqs. (88), which are due to velocity-dependent interactions in the Lagrangian, were obtained first by Helmholtz[30] and easily can be obtained by the standard method of Lagrangian dynamics.

The conservation laws (79) and (80), or equivalently Eqs. (87) were obtained above as special cases of the general conservation laws (66) and (67), using the retarded (advanced) Green's functions (64) with (76). They also could have been obtained directly from the equations of motion (60) with (61), using (64) with (76), which then depend only on the one time, say, $t_1$. Therefore, these equations can be treated by standard methods, and the solutions are specified by 6n independent data, the initial positions and velocities as in Newtonian mechanics, except that the initial data are not specified at a single time, but those of the i-th particle at $t_i^*$ connected by (77).

We also note that the addition of condition (76) to the time-symmetric Green's function (65) instead of the asymmetric one (64) does *not* lead either to conservation laws or to equations of motion which depend only on n times. Whether these equations have solutions which nevertheless depend only on 6n independent initial data is an open question, just as for the analogous equations of special relativistic dynamics

## 5. An example of generalized Galilei-invariant mechanics

General Lagrange equations of motion (60) with forces (61) were given in Section III. Here, the Green's function is chosen to be a special case of Eq. (64), viz.,

$$4\pi G_{ij} = \delta(t_{ij} - T_i + T_j) V_{ij}(S_{ij}^2),$$
(90)

where $\delta(t_{ij} - T_i + T_j)$ is the Dirac "delta function," $V_{ij}$ is the two-body potential energy, and, as shown in Sec. 4 using (70), (76), and (77), this is the case that has an initial value problem similar to Newton's. For brevity in Eqs. (91) through (95) below, $T_{ij}$ sometimes replaces the constants $T_i - T_j$ where confusion should not result. This delta function has another advantage: the one-body potential energy $U_i$ defined in Eq. (59) first can be integrated and then the equations of motion for ρ = 1, 2, 3 calculated by (60), or (90) can be substituted into (61) which can be used in (60) for all ρ, or (90) can be substituted in (55), one of the two integrations carried out, and then use the resulting 3-d Lagrangian to obtain the equations of motion. For the spatial components, of course, the equations are the same and result in

$$\begin{aligned} m_i \vec{a}_i = &-\sum_{j>i} \left\{ (V_{ij}') \left[ 2\vec{S}_{ij} + (t_{ij}) \left( \frac{d\vec{S}_{ij}}{d\tau_i} + \frac{d\vec{S}_{ij}}{d\tau_j} \right) \right] + (V_{ij}'') \left[ (t_{ij})(\vec{S}_{ij}) \left( \frac{dS_{ij}^2}{d\tau_i} + \frac{dS_{ij}^2}{d\tau_j} \right) \right] \right\} \Big|_{t_j = t_i - T_{ij}} \\ &+ \sum_{j<i} \left\{ (V_{ji}') \left[ 2\vec{S}_{ji} - (t_{ji}) \left( \frac{d\vec{S}_{ji}}{d\tau_i} + \frac{d\vec{S}_{ji}}{d\tau_j} \right) \right] - (V_{ji}'') \left[ (t_{ji})(\vec{S}_{ji}) \left( \frac{dS_{ji}^2}{d\tau_i} + \frac{dS_{ji}^2}{d\tau_j} \right) \right] \right\} \Big|_{t_j = t_i + T_{ji}}, \end{aligned}$$
(91)

where a prime on the potential energy means a derivative with respect to its argument. Eq. (91) is explicitly Galilei-invariant. Replacement of $\vec{S}_{ij}$ by its definition Eq. (29) in (91) yields finally

$$m_i \vec{a}_i = \sum_{j>i} \{(V'_{ij})\left[-2\vec{r}_{ij} + 2\vec{v}_j T_{ij} + \tfrac{1}{2} T_{ij}^2 (\vec{a}_i + \vec{a}_j)\right]$$

$$+ (V''_{ij}) T_{ij} \left[-\vec{r}_{ij} + \tfrac{1}{2}(T_{ij})(\vec{v}_i + \vec{v}_j)\right] \frac{dS^2_{ij}(t_i, t_i - T_{ij})}{dt_i} \} \big|_{t_j = t_i - T_{ij}}$$

$$+ \sum_{j<i} \{(V'_{ji})\left[2\vec{r}_{ji} - 2\vec{v}_j T_{ji} + \tfrac{1}{2} T_{ji}^2 (\vec{a}_i + \vec{a}_j)\right]$$

$$+ (V''_{ji}) T_{ji} \left[-\vec{r}_{ji} + \tfrac{1}{2}(T_{ji})(\vec{v}_i + \vec{v}_j)\right] \frac{dS^2_{ji}(t_i + T_{ji}, t_i)}{dt_i} \} \big|_{t_j = t_i + T_{ji}}, \quad (92)$$

where the derivative of the invariant is

$$\frac{dS^2_{ij}(t_i, t_i - T_{ij})}{dt_i} = 2\vec{v}_{ij} \cdot \vec{r}_{ij} - (\vec{v}_i^{\,2} - \vec{v}_j^{\,2}) T_{ij} - (\vec{a}_i + \vec{a}_j) \cdot \vec{r}_{ij} T_{ij} + \tfrac{1}{2}(\vec{a}_i + \vec{a}_j) \cdot (\vec{v}_i + \vec{v}_j) T_{ij}^2.$$

When $T_{ij} = 0$, Eq. (92) reduces to Newton's mechanics. For the sake of brevity, most arguments in (92) are not exhibited, but they should be obvious. Eq. (92) involves not only the Newtonian "forces" $V'_{ij}$, but also their derivatives $V''_{ij}$. For just two particles, the two equations of motion follow from Eq. (92) as

$$m_1 \vec{a}_1 = \{(V'_{12})[-2\vec{r}_{12} + 2\vec{v}_2 T_{12} + \tfrac{1}{2}(\vec{a}_1 + \vec{a}_2) T_{12}^2]$$

$$+ (V''_{12}) T_{12}[-\vec{r}_{12} + \tfrac{1}{2}(\vec{v}_1 + \vec{v}_2) T_{12}] (\frac{dS^2_{12}(t_1, t_1 - T_{12})}{dt_1})\} \big|_{t_2 = t_1 - T_{12}} \equiv \vec{F}_{12}(t_1, t_1 - T_{12}),$$

$$(93)$$

$$m_2 \vec{a}_2 = \{(V'_{12})[2\vec{r}_{12} - 2\vec{v}_1 T_{12} + \tfrac{1}{2}(\vec{a}_1 + \vec{a}_2) T_{12}^2]$$

$$+ (V''_{12}) T_{12}[-\vec{r}_{12} + \tfrac{1}{2}(\vec{v}_1 + \vec{v}_2) T_{12}] (\frac{dS^2_{12}(t_2 + T_{12}, t_2)}{dt_2})\} \big|_{t_1 = t_2 + T_{12}} \equiv \vec{F}_{21}(t_2, t_2 + T_{12}).$$

A trivial change of variable $t_2 = t_1 - T_{12}$ in either of Eqs. (93) makes both equations have the same arguments. Since the two times $t_1$ and $t_2 = t_1-T_{12}$ differ only by a constant, differentiation of an expression that depends on both with respect to either one is like a total time derivative in Newtonian mechanics. Thus, in Eqs. (93), it is true that

$$\frac{dS^2_{12}(t_1, t_1 - T_{12})}{dt_1} = \frac{dS^2_{12}(t_2 + T_{12}, t_2)}{dt_2} \big|_{t_2 = t_1 - T_{12}}.$$

Except when $T_{12} = 0$, the forces $\vec{F}_{12}$ and $\vec{F}_{21}$ in (93) do not obey Newton's third law of motion, not even a generalized version. However, their sum is a total time derivative, giving an alternative way to calculate the conserved momentum expression which is shown in (95) below. A transformation to an inertial frame where the sum of the velocities is zero simplifies these equations, but does not relieve the velocity dependence.

Obtaining the equations of motion by substitution of (90) into (55), performing one of the two integrations, reading off the 3-d Lagrangian and then using it in Lagrange's equations gives a form equivalent to (93) above

$$(d/dt_1)\{m_1 \vec{v}_1 - V'_{12}[-\vec{r}_{12} T_{12} + \tfrac{1}{2}(\vec{v}_1 + \vec{v}_2) T_{12}^2]\} = (V'_{12})[-2\vec{r}_{12} + (\vec{v}_1 + \vec{v}_2) T_{12}],$$

$$(94)$$

$$(d/dt_2)\{m_2 \vec{v}_2 - V'_{12}[-\vec{r}_{12} T_{12} + \tfrac{1}{2}(\vec{v}_1 + \vec{v}_2) T_{12}^2]\} = (V'_{12})[2\vec{r}_{12} - (\vec{v}_1 + \vec{v}_2) T_{12}],$$

which exhibits the canonical momenta and gives forces that do obey a generalized Newton's third law.

The ten conserved quantities written in terms of derivatives of the two-body potential energies were shown in Eqs. (87); choosing $V_{ij}(S_{ij}^2)$ for two particles gives

$$E(t_1^*, t_1^* - T_{12}) = \tfrac{1}{2} m_1 \vec{v}_1^2 + \tfrac{1}{2} m_2 \vec{v}_2^2 + V_{12} + (V_{12}')T_{12}[(\vec{v}_1 + \vec{v}_2)\cdot\vec{r}_{12} - \tfrac{1}{2}(\vec{v}_1 + \vec{v}_2)^2 T_{12}],$$
$$\vec{P}(t_1^*, t_1^* - T_{12}) = m_1 \vec{v}_1 + m_2 \vec{v}_2 + (V_{12}')T_{12}[2\vec{r}_{12} - (\vec{v}_1 + \vec{v}_2)T_{12}],$$
$$\vec{G}(t_1^*, t_1^* - T_{12}) = (m_1 \vec{r}_1 - t_1^* m_1 \vec{v}_1) + (m_2 \vec{r}_2 - t_2^* m_2 \vec{v}_2) + (V_{12}')[-\vec{r}_{12} - \tfrac{1}{2}(\vec{v}_1 + \vec{v}_2)T_{12}](t_1^* + t_2^*)T_{12},$$
$$\vec{L}(t_1^*, t_1^* - T_{12}) = \vec{r}_1 \times m_1 \vec{v}_1 + \vec{r}_2 \times m_2 \vec{v}_2 + (V_{12}')T_{12}[-2\vec{r}_1 \times \vec{r}_2 - \tfrac{1}{2}(\vec{r}_1 + \vec{r}_2) \times (\vec{v}_1 + \vec{v}_2)T_{12}].$$
(95)

To this point, the potential energy function is unspecified except for its argument. For the presumably interesting case of a generalized Newtonian-like gravity with $\vec{r}_{12}(t)$ replaced by $\vec{S}_{12}(t_1,t_2)$, the generalized gravitational potential energy and its derivatives are

$$V_{12}(S_{12}^2) = -Gm_1 m_2 (S_{12}^2)^{-\tfrac{1}{2}};\quad V_{12}' = \tfrac{1}{2} Gm_1 m_2 (S_{12}^2)^{-\tfrac{3}{2}};\quad V_{12}'' = -\tfrac{3}{4} Gm_1 m_2 (S_{12}^2)^{-\tfrac{5}{2}}.$$
(96)

All of Eqs. (92)-(96) reduce to Newtonian ones when the time delay $T_{ij}$ is set to zero.

As noted in the introduction, use of (96) in (93) does not lead to motions familiar from Newtonian physics. This is easiest seen by looking at the conserved momentum (95). For equal masses with equal and opposite velocities *at all times* $t_1$ and $t_1 - T_{12}$ (which implies that the accelerations are also equal and opposite at those times) the conserved momentum reduces to

$$\vec{P}(t_1^*, t_1^* - T_{12}) = (V_{12}')T_{12} 2\vec{r}_{12},$$
(97)

which clearly cannot be constant in both magnitude and direction for either the case of equal mass particles approaching one another on a straight line or on different parallel lines situated on a line perpendicular to both velocities which, in Newtonian mechanics, would be expected to result in a circle. Unequal accelerations follow from the ratio of Eqs. (93), using (96).

## 6.    Discussion

As an alternative to slow motion descriptions of point particles[1, 6, 7, 8] following from GRT, we have suggested use of the four-dimensional classical mechanics created by the second author[13] to be invariant under the transformations of the proper orthochronous inhomogeneous Galilei group in a generalized Lagrangian form Eqs. (60) with (61). In order to construct this mechanics, the two-body invariants of the proper orthochronous inhomogeneous Galilei group were derived here using, for the first time[21] in 4-d, a *non-singular* two-index tensor given both in (3+1) form (27) and a new explicit tensor form (36); the two-body invariants trivially were modified to be the invariants of the full group. The usual ten conserved quantities were presented, each involving integrals over the world lines as in special relativistic theories. Eqs. (66) and (67) follow from Noether's theorem applied to the variational principle (55) with (57) and (58). For the special Lagrange case of a time-asymmetric retarded (advanced) interaction Eq. (64) involving a correlated interlocking set of events (75) ensured by (76) with one event per world line, we were able to show that terms involving integrals over world lines vanished in the ten conserved quantities when the times of evaluation were chosen to satisfy (77). Thus, the conserved quantities for the time-asymmetric interaction Eq. (64) involve only algebraic forms as in Newtonian mechanics and this mechanics has a Newtonian-like initial value problem.

One example of this mechanics was given. It involves a time-asymmetric interaction (90) whose two-body potential energy is a function only of the invariant $S_{ij}^2$. The general equations of motion were given in (91) or (92); for two particles the equations of motion were exhibited in (93) and the algebraic form conserved quantities in (95). For generalized Newtonian gravity the potential energy and its derivatives were presented in (96). Inferences on expected motions restricted by this form of mechanics were given in the paragraph surrounding (97).

The fact that we chose $S_{ij}^2$ as the argument of $V_{ij}$ in (90) should not mislead the reader into thinking that it is the only possibility. If we demand that the Galilei-invariant argument has dimensions of distance (or distance squared) in order to have an easy connection to the well-known Newtonian case when $T_{ij} = 0$, there are several possibilities: $S_{ij}^2$, $v_{ij}^2 t_{ij}^2$, $K_{ij} t_{ij}$, and linear combinations thereof. However, none lead to better insights than our obvious choice in (90).

The equations of motion found in this paper cannot be verified in the laboratory, because the time delay of the interactions is expected, as a first approximation on the basis of the special theory of relativity to be the average interparticle separation divided by the speed of light. Thus, the natural application (if any) of this theory is to astrophysical problems. But, that would take us well beyond the objective of this paper which is to bring this form of classical mechanics to the attention of the physics community. Certainly, an instantaneous Newtonian interaction seems not at all realistic when dealing with astrophysical applications.

To apply the Lagrangian equations of motion to the solar system, one might try an expansion in powers of the time delay. However, this results in a precisely zero result to first order. The second order equations are quite tedious.

We deliberately neglected the time-symmetric Green's function (65). Although it can be connected to GRT via a limit of the time-symmetric Havas-Goldberg[5] variational principle, as pointed out in Sec. 4, this case does not have vanishing double integrals in the multiple-time conserved quantities and consequently, does not have a Newtonian-like initial value problem. To our knowledge, only the time-asymmetric interaction (64) has this valuable property.

**Acknowledgments**

A Sabbatical Leave for the fall semester of 2002 granted by Philadelphia University supported the first author and he thanks Temple University for allowing him to spend it in residence while some of this work was completed.

**Appendix 1: Invariants involving accelerations**

The invariants involving only $s_{ij}^\mu$, $S_{ij}^\mu$, and $v_{ij}^\mu$ were given in Sec. II in Eqs. (30)-(33). A complete list including the above and those involving accelerations are given here. All are calculated using the explicit representation Eq. (27) of $k_{ij\mu\nu}$ and the accelerations $a_i^\nu$ defined in (12) as well as Eqs. (24) and (29). Using the symmetry in the indices $\mu$ and $\nu$ reduces the number that must be displayed. First, contracting $k_{ij\mu\nu} S_{ij}^\mu$ with $S_{ij}^\nu, v_i^\nu, v_j^\nu, v_{ij}^\nu, a_i^\nu, a_j^\nu$, and $a_{ij}^\nu$ respectively yields

$$\vec{S}_{ij}^2, \ \tfrac{1}{2}\vec{v}_{ij}\cdot\vec{S}_{ij}, \ -\tfrac{1}{2}\vec{v}_{ij}\cdot\vec{S}_{ij}, \ \vec{v}_{ij}\cdot\vec{S}_{ij}, \ \vec{a}_i\cdot\vec{S}_{ij}, \ \vec{a}_j\cdot\vec{S}_{ij}, \ \vec{a}_{ij}\cdot\vec{S}_{ij}. \tag{A1}$$

Contracting $k_{ij\mu\nu} v_i^\mu$ with $v_i^\nu, v_j^\nu, v_{ij}^\nu, a_i^\nu, a_j^\nu$, and $a_{ij}^\nu$ respectively gives

$$\tfrac{1}{2}\vec{v}_{ij}^2, \ 0, \ \tfrac{1}{2}\vec{v}_{ij}^2, \ \tfrac{1}{2}\vec{a}_i\cdot\vec{v}_{ij}, \ \tfrac{1}{2}\vec{a}_j\cdot\vec{v}_{ij}, \ \tfrac{1}{2}\vec{a}_{ij}\cdot\vec{v}_{ij}, \tag{A-2}$$

while contracting $k_{ij\mu\nu} v_j^\mu$ with $v_j^\nu, v_{ij}^\nu, a_i^\nu, a_j^\nu$, and $a_{ij}^\nu$ results in

$$\tfrac{1}{2}\vec{v}_{ij}^2, \ -\tfrac{1}{2}\vec{v}_{ij}^2, \ -\tfrac{1}{2}\vec{a}_i\cdot\vec{v}_{ij}, \ -\tfrac{1}{2}\vec{a}_j\cdot\vec{v}_{ij}, \ -\tfrac{1}{2}\vec{a}_{ij}\cdot\vec{v}_{ij}. \tag{A-3}$$

Keeping to the pattern, contraction of $k_{ij\mu\nu} v_{ij}^\mu$ with $v_{ij}^\nu, a_i^\nu, a_j^\nu$, and $a_{ij}^\nu$ leads to

$$\vec{v}_{ij}^2, \ \vec{a}_i\cdot\vec{v}_{ij}, \ \vec{a}_j\cdot\vec{v}_{ij}, \ \vec{a}_{ij}\cdot\vec{v}_{ij}. \tag{A-4}$$

Finally, contraction of $k_{ij\mu\nu} a_i^\mu$ with $a_i^\nu, a_j^\nu$, and $a_{ij}^\nu$ gives

$$\vec{a}_i^2, \ \vec{a}_i\cdot\vec{a}_j, \ \vec{a}_i\cdot\vec{a}_{ij}, \tag{A-5}$$

while contracting $k_{ij\mu\nu} a_j^\mu$ with $a_j^\nu$, and $a_{ij}^\nu$ yields

$$\vec{a}_j^2, \ \vec{a}_j \cdot \vec{a}_{ij}, \tag{A-6}$$

and contracting $k_{ij\mu\nu}a_{ij}^\mu$ with $a_{ij}^\nu$ results in

$$\vec{a}_{ij} \cdot \vec{a}_{ij}. \tag{A-7}$$